\documentclass[aps,prl,twocolumn,groupedaddress,showpacs]{revtex4}

\usepackage{graphicx}
\usepackage{amsmath}
\usepackage{amssymb}
\usepackage{bm}

\begin{document}
\title{Adiabatic loading of a Bose-Einstein condensate in a 3D optical lattice}
\author{T.\ Gericke, F.\ Gerbier, A.\ Widera, S.\ F{\"o}lling, O.\ Mandel, and I.\ Bloch}
\affiliation{Institut f{\"u}r Physik, Johannes
Gutenberg-Universit{\"a}t, 55099 Mainz, Germany.}
\date{\today}
\begin{abstract}
We experimentally investigate
the adiabatic loading of a
Bose-Einstein condensate into an optical lattice potential. The
generation of excitations during the ramp is detected by a
corresponding decrease in the visibility of the interference
pattern observed after free expansion of the cloud. We focus on
the superfluid regime, where we show that the limiting time scale
is related to the redistribution of atoms across the lattice by
single-particle tunneling.
\end{abstract}
 \maketitle
%
%
%
%
The observation of the superfluid to Mott insulator (MI)
transition \cite{greiner2002a} undergone by an ultracold Bose gas
in an optical lattice has triggered a lot of experimental and
theoretical activities (see
\cite{zwerger2003a,bloch2005a,jaksch2005a}). This system allows to
experimentally produce strongly-correlated quantum systems in a
well controlled environment,  with applications ranging from
the realization of novel quantum phases in multi-component systems
(see, e.g., \cite{jaksch2005a} and references therein) to the
implementation of collisional quantum gates
\cite{jaksch1999a,mandel2003a,mandel2003b}.

Most of these proposals rely on producing a system very close to
its ground state. However, in experiments so far, the
successful production of an ultracold gas in the optical lattice
relies on adiabatic transfer. The condensate is first produced
and evaporatively cooled in order to minimize the thermal
fraction. The almost pure condensate is subsequently transferred
into the lattice potential by ramping up the laser intensities as
slow as possible in order to approach the adiabatic limit. For too
fast a ramp, excitations are generated in the system, which
eventually results in heating after the cloud has equilibrated at
the final lattice depth.

It is important in this respect to note the existence of two
energy scales in this problem, related to the single-particle band
structure on the one hand or to the many-body physics within the
lowest Bloch band on the other hand. Adiabaticity with respect to
the band structure is associated with the absence of interband
transitions. The characteristic time scales are on the order of
the inverse recoil frequency, typically hundreds of microseconds.
Experimentally, adiabaticity with respect to the band structure is
easily ensured, and can be checked by detecting atoms in the
higher Bloch bands \cite{greiner2001a,denschlag2002a}.
Adiabaticity with respect to the many-body dynamics of the system
involves considerably longer time scales (on the order of tens or
even hundreds of milliseconds). Theoretical studies of the
loading dynamics have been reported in
\cite{sklarz2002a,sklarz2002b,clark2004a,mackagan2005a,zakrzewski2005a,isella2005a}.
It is clear that non-adiabatic effects are more pronounced in the
superfluid phase. The insulator phase is indeed expected to be
quite insensitive to such effects, due to the presence of an
energy gap. In this paper, we focus on the superfluid phase. Our
goal is to to clarify experimentally how slowly the loading has to
proceed to minimize unwanted excitations and heating.

To this aim, we make use of long range phase properties
characteristic to a BEC. A key observable for ultracold Bose gases
in optical lattices is the interference pattern observed after
releasing the gas from the lattice and letting it expand for a
certain time of flight
\cite{orzel2001a,greiner2001a,greiner2002a,stoeferle2004a,gerbier2005a,gerbier2005b}.
The contrast of this interference pattern is close to unity when
most atoms belong to the condensate, but diminishes with the
condensed fraction as the cloud temperature increases. Hence, the
visibility of this interference pattern can be used to investigate
the dynamical loading of a condensate into the optical lattice. We
focus here on a specific lattice depth in the superfluid regime.
We find a characteristic time scale of $\sim 100\,$ms above which
the visibility of the interference pattern appears to be
stationary. We show how it relates to the redistribution of atoms
in the lattice through single-particle tunneling.

\begin{figure}
\includegraphics[width=8.6cm]{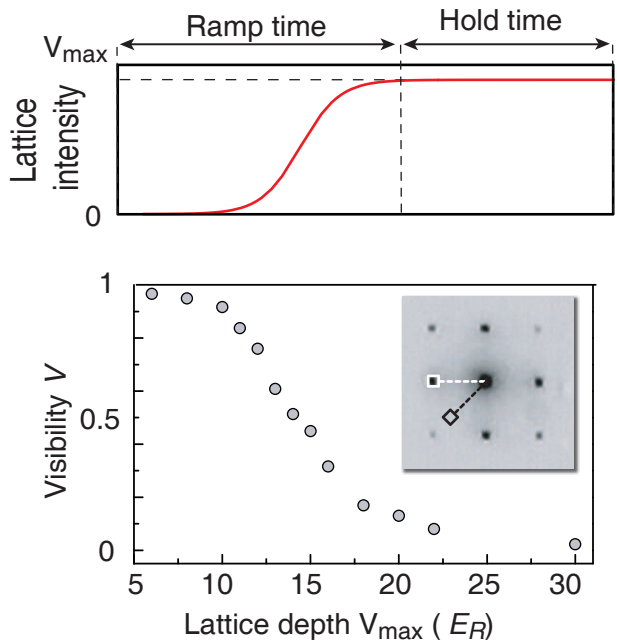} \caption{({\bf a})Sketch of the
time profile used to ramp up the lattice depth to its maximum
value $V_{\rm max}$. ({\bf b}) Evolution of the visibility of the
interference pattern as the lattice depth is increased. The set of
data shown corresponds to $\sim 3\times10^5$ atoms (grey
circles).} \label{Fig1}
\end{figure}

In our experiment, a $^{87}$Rb Bose-Einstein condensate is loaded
into an optical lattice created by three orthogonal pairs of
counter-propagating laser beams (see \cite{greiner2002a} for more
details). The superposition of the lattice beams, derived from a
common source at a wavelength $\lambda_L=850~$nm, results in a
simple cubic periodic potential with a lattice spacing
$d=\lambda_L/2=425~$nm. The lattice depth $V_0$ is controlled by
the laser intensities, and is measured here in units of the
single-photon recoil energy, $E_r=h^2/2m\lambda_L^2\approx
h\times3.2~$kHz. The optical lattice is ramped up in a time
$\tau_{\rm ramp}$, using a smooth waveform that minimizes sudden
changes at both ends of the ramp. The ramp form is calculated
numerically. The program imposes that the lattice depth is
initially zero and reaches its final value after a time $\tau_{\rm ramp}$. In between (for times $0\leq t \leq\tau_{\rm ramp}$), it
tries to match the functional form
\begin{equation}\label{Vt}
V_0(t)=\frac{V_{\rm max}}{1+\exp\left(-\alpha\frac{t}{\tau_{\rm
ramp}}\right)},
\end{equation}
with $\alpha=20$, by a piecewise linear approximation. The
function (\ref{Vt}) has been proposed in \cite{clark2004a} and is
shown in Fig. \ref{Fig1}a. After this ramp, the cloud is held in
the lattice potential for a variable hold time $t_{\rm hold}$ (see
Fig. \ref{Fig1}a), during which the system can
re-thermalize. After switching off the optical and magnetic
potentials simultaneously and allowing for typically $t=10-22~$ms
of free expansion, standard absorption imaging of the atom cloud
yields a two-dimensional map of the density distribution $n$
(integrated along the probe line of sight).

To extract quantitative information from time-of-flight pictures,
we use the usual definition of the visibility of interference
fringes,
\begin{equation}\label{V}
\mathcal{V}=\frac{n_{\rm max}-n_{\rm min}}{n_{\rm max}+n_{\rm
min}}.
\end{equation}
In this work, we follow the method introduced in
\cite{gerbier2005a} and measure the maximum density $n_{\rm max}$
at the first lateral peaks of the interference pattern, ({\it
i.e.} at the center of the second Brillouin zone). The minimum
density $n_{\rm min}$ is measured along a diagonal with the same
distance from the central peak (see inset in Fig.~\ref{Fig1}b). In
this way, the Wannier envelope is the same for each term and
cancels out in the division (see \cite{gerbier2005a}). For a given
two-dimensional absorption image, four such pairs exist and the
corresponding values are averaged to yield the visibility. In
addition, it is worth pointing out that this is an essentially
model-independent characterization of the many-body system.

\begin{figure}
\includegraphics[width=8.6cm]{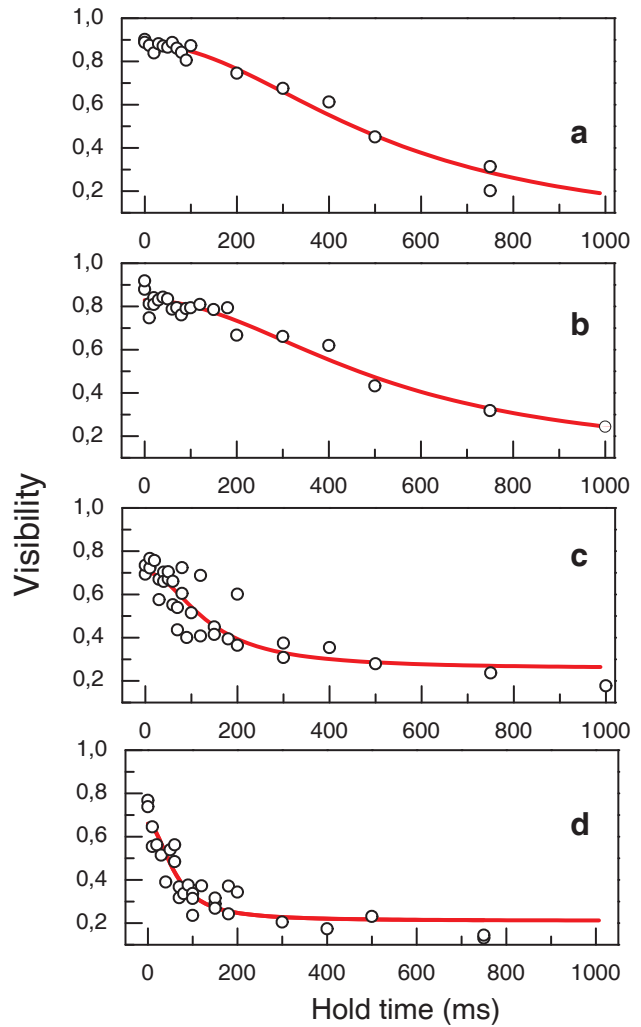} \caption{Time evolution of the visibility for a fixed final
depth $V_{\rm max}=10~E_r$ and various ramp times: ({\bf a}):
$\tau_{\rm ramp}=160~$ms, ({\bf b}): $\tau_{\rm ramp}=80~$ms,
({\bf c}): $\tau_{\rm ramp}=40~$ms, and ({\bf d}): $\tau_{\rm
ramp}=20~$ms.}\label{Fig2}
\end{figure}
For comparison, we reproduce here measurements of the visibility
as a function of lattice depth, similar to those presented in
\cite{gerbier2005a}. The data corresponds to a given total atom
number $N\approx 3\times10^5$ atoms (grey circles in
Fig.~\ref{Fig1}b). For lattice depths larger than $12.5~E_r$, the
system is in the insulating phase, as demonstrated in
\cite{greiner2002a}. Yet, the visibility remains finite well above
this point. For example, at a lattice depth of $15~E_r$, the
contrast is still around $30 \%$, reducing to a few percent level
only for a rather high lattice depth of $30~E_r$. In
\cite{gerbier2005a}, we have shown that such a slow loss in
visibility is expected in fact even in the ground state
of the system, being a consequence of the admixture of
particle/hole pairs to the ground state for finite lattice depths.
For these experiments, a ramp time of 160 ms followed by a hold
time of 40 ms were used. We will show later on that this is slow
enough to ensure adiabaticity.

We now focus on the superfluid regime, at a lattice depth
$V_0=10~E_r$. To investigate how the system is affected by the
ramping procedure, we fix the ramp time $\tau_{\rm ramp}$ and
record how the interference pattern evolves as a function of
hold time $\tau_{\rm hold}$. Examples of such measurements are
shown in Fig.~\ref{Fig2}a-d.

For the slowest ramp shown, $\tau_{\rm ramp}=160$\,ms, the
visibility decreases with a time constant $\sim 500$\,ms. In fact,
a very similar behavior is observed as soon as the ramp time
exceeds $\tau_{\rm ramp}>100$\,ms. Since this behavior is
essentially independent of the ramp speed, it points to the
presence of heating mechanisms which significantly degrade the
visibility on a time scale of several hundred ms. The source of
heating could be technical, due for instance to intensity or
pointing fluctuations of the lattice beams, or intrinsic
processes. An unavoidable process is for instance atomic
spontaneous emission following excitations by one of the lattice
beams. Here we attempt a crude estimate of the effect of
spontaneous emission on visibility as follows. The heating rate
$\Gamma_{\rm heat}\sim \Gamma_{\rm sp}\times(h^2/2 m \lambda_0^2)$
is given by the rate $\Gamma_{\rm sp}$ at which such events happen
due to {\it all} lattice beams, times the recoil energy $h^2/2 m
\lambda_0^2$, where $\lambda_0\approx 780$\,nm
is the resonant wavelength of the D$_2$ transition. To
estimate the time scale $T_{\mathcal{V}}$ over which the
visibility vanishes, we compute the time over which the energy
gain per particle, $\Gamma_{\rm heat}T_{\mathcal{V}}$, is on the
order of the critical temperature $T_c$ times Boltzmann's constant
$k_B$. In a lattice potential with approximately one atom per
site, we have $k_B T_c \sim z J$, where $J$ is the tunneling
matrix element and $z=6$ the number of nearest neighbors in a
three dimensional cubic lattice. This yields a final estimate
\begin{equation}
T_{\mathcal{V}} \sim \Gamma_{\rm sp}^{-1} \frac{z J}{E_r}.
\end{equation}
For a lattice depth of $10~E_r$, we calculate a tunneling energy
$z J/E_r\sim0.12$ and a total scattering rate $\Gamma_{\rm
sp}\sim0.2$\,$s^{-1}$, giving $T_{\mathcal{V}}\sim 0.6$\,s, in
good agreement with the observed time scale. We conclude that on
the time scales considered here, photon scattering contributes
significantly to the observed heating rate. Although heating due
to technical noise is not excluded, the bound above implies that
it is not much more severe than photon scattering. We note that in
principle, the decay in visibility could be used to measure the
heating rate, provided the dependance of the visibility on
temperature is known.

\begin{figure}
\includegraphics[width=8cm]{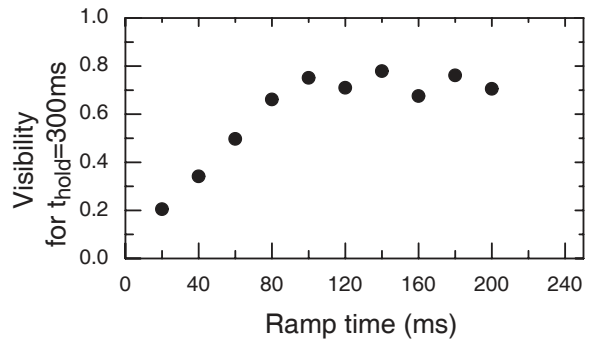}
\caption{Measured visibility versus ramp time, for a fixed 300 ms
hold time.} \label{Fig3}
\end{figure}

\begin{figure}
\includegraphics[width=8cm]{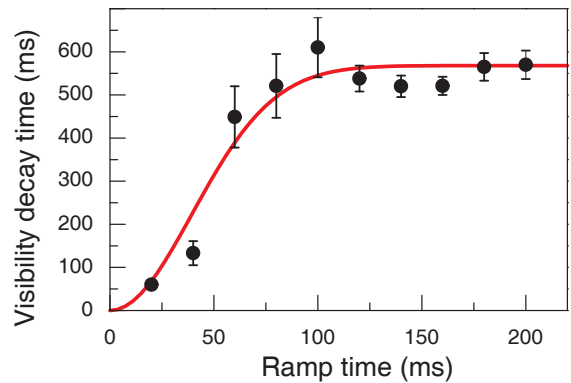}
\caption{Measured visibility decay time versus ramp time. The time
constant is the half-width at half-maximum of a Lorentzian fit to
the data. Error bars indicate the statistical error. The solid
line is a fit to an inverted Gaussian, returning a time scale of
$\sim 60$\,ms for the visibility decay time to become independent
of the ramp.} \label{Fig4}
\end{figure}

When the ramp times is decreased below $\tau_{\rm ramp}=100$\,ms,
the visibility decreases in a qualitatively different way. As
shown in Fig.~\ref{Fig2}b-d, the decay occurs with a much shorter
time constant. Note that at long hold times $\tau_{\rm
hold}=800$\, ms, the visibility has dropped to $\mathcal{V}\sim
0.2$, a value almost independent of the ramp time. This is
consistent with our earlier claim that heating dominates at long
times. We attribute the short-times decay to the generation of
excitations by a finite ramp speed. To compare the different ramp
times, we plot in Fig.~\ref{Fig3} the measured visibility as a
function of ramp time for a fixed hold time of 300 ms, long enough
for the excitations generated during the ramp to relax, but short
enough that heating effects do not blur all differences between
the different ramps. A time scale of $\sim 100$\,ms clearly
emerges, above which the ramp time can be increased without
noticeable effect. To check whether this could be an artifact of
our plotting method, we have fitted the data to an empirically
chosen Lorentzian form. The half-width at half-maximum of the
Lorentzian was taken as the visibility decay time. As seen from
Fig.~\ref{Fig4}, the same behavior is found, with a characteristic
$1/e$ ramp time $\approx 60$\,ms.

\begin{figure}
\includegraphics[width=8cm]{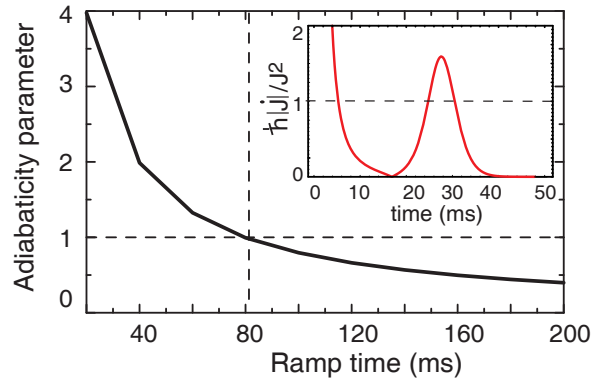}
\caption{Adiabaticity parameter (see text) plotted versus ramp
time. In the calculation, we have used the actual ramp form as
generated in the experiment to describe the lattice depth
increase. In the inset, we show how the adiabaticity parameter changes
in time as the lattice depth is ramped up. A ramp time of
$\tau_{\rm ramp}=50$\,ms has been used for this plot. }
\label{Fig5}
\end{figure}

This time scale can be understood using elementary arguments.
Adiabatic evolution in a quantum system with time-dependent
hamiltonian requires the condition
\begin{equation}\label{adiab}
|\dot{H}| \ll \hbar |\omega_{fi}|^2,
\end{equation}
where $\dot{H}$ is the derivative of the hamiltonian and where
$\omega_{fi}$ is the Bohr frequency of transition between the
(instantaneous) eigenstates $|i\rangle$ and $|f\rangle$.
Eq.~(\ref{adiab}) should be fulfilled at all times. For an
ultracold gas in an optical lattice, three energy scales appear:
(i) the tunneling matrix element $J$ defining the rate of hopping
from site to site, (ii) the on-site interaction energy $U$ and,
(iii) the energy associated to the ``external'' confinement
potential usually present on top of the lattice. This potential
$V_{\rm ext}$ results from the combination of the magnetic
trapping potential in which the condensate is initially produced,
and of an additional confinement due to the Gaussian shape of the
lasers producing the optical lattice. In practice, $V_{\rm ext}$
is nearly harmonic, with a trapping frequency $\Omega=\omega_{\rm
ext}\approx\sqrt{\omega_{\rm m}^2+8V_0/mw^2}$, where $\omega_{\rm
m}=2\pi\times16$\,Hz is the oscillation frequency in the magnetic
trap and where $w\approx136\,\mu$m is the laser beam size. With a
proper choice of the laser beam sizes, the increase in $U$ and
$\Omega$ are such that the equilibrium size of the condensate as
calculated in the Thomas-Fermi approximation varies little
during the ramp \cite{PhDGreiner}. In addition, when the lattice
depth is increased, $J$ drops exponentially fast whereas $U$ and
$\Omega$ increase slowly. Hence, the question of adiabaticity
essentially reduces to whether the atoms can redistribute through
tunneling in order to adapt the size of the system to the
instantaneous Thomas-Fermi shape. This suggests to take
$|\dot{H}|\sim |\dot{J}|$ and $\omega_{fi}\sim J / \hbar$ in
Eq.~(\ref{adiab}), giving the following definition for the
adiabaticity parameter $\mathcal{A}$,
\begin {equation}\label{A }
\mathcal{A} =\underset{0 \leq t \leq \tau_{\rm ramp}}{\rm
Max}\left[\frac{\hbar |\dot{J}|}{ J^2}\right].
\end {equation}
For a given ramp, we require $\mathcal{A}\ll 1$ to ensure an
adiabatic loading of the cloud into the lattice.  In the inset of
Fig.~\ref{Fig5}, we show the quantity $\hbar |\dot{J}|/J^2$
calculated for $\tau_{\rm ramp}= 50$\, ms. We have calculated this
curve for a final lattice depth of $10 E_R$ using the lattice ramp
given in Eq.~(\ref{Vt}), and the analytical estimate
\cite{zwerger2003a},
\begin{equation}\label{J}
\frac{J}{E_r}\approx\sqrt{\frac{8}{\pi}}
(V_0/E_r)^{3/4}\exp(-2\sqrt{V_0/E_r}).
\end{equation}
The sharp decrease for short times in the inset of Fig.~\ref{Fig5}
follows from the inadequacy of the tight-binding approximation
under which Eq.~(\ref{J}) is derived. We ignore this feature, and
calculate $\mathcal{A}$ from the peak value occurring near
$t\approx\tau_{\rm ramp}/2$, where the rate of change of the
lattice depth is highest. We have repeated the calculation for
several ramp times (see Fig.~\ref{Fig5}). We find that the
critical $\mathcal{A}=1$ corresponds to a ramp time $\tau_{\rm
ramp}\approx 80$\, ms, close to the experimental findings. This
suggests that for our experimental parameters, the loading is
indeed limited by single-particle tunneling.

In conclusion, we have studied how the visibility of the
interference pattern was affected by the speed at which the
lattice was ramped up. A time scale of $\sim 100$ ms was found for
adiabatic loading in the optical lattice . In this paper, we
focused on the dynamics properties in the superfluid regime. Even
more interesting is the dynamical evolution of the system as the
superfluid-to-Mott-insulator transition is crossed. An important
and still open question is in particular how reversible this
transition is. Experiments
\cite{orzel2001a,greiner2002a,stoeferle2004a,xu2005a} found that
one could ramp up the lattice intensity and reach the regime where
phase coherence is lost, then ramp down the lattice and regain it
back. To what extent the initial phase coherent state can be
recovered, and what are the limiting mechanisms has however not
been studied. Also, the recent availability of virtually exact
methods for one-dimensional systems \cite{clark2004a} could allow
for a precise comparison with the experimental data, in a strongly
non-equilibrium regime where theory 
is still in progression.
Finally, new dynamical effects are predicted, such as oscillation
of the order parameter and vortices-driven relaxation analog to
the Kibble-Zurek mechanism \cite{altman2002a}.

Our work is supported by the Deutsche Forschungsgemeinschaft
(SPP1116), AFOSR and the European Union under a Marie-Curie
Excellence grant (OLAQUI). FG acknowledges support from a
Marie-Curie Fellowship of the European Union.
%
%
\bibliography{mott}
\bibliographystyle{apsrev}


\end{document}